\documentclass[12pt]{article}

\usepackage{latexsym}
\usepackage{amsmath,,calrsfs}
\usepackage{amsfonts}
\usepackage{amssymb}
\usepackage{amscd}
\usepackage{fancybox}
\usepackage{cite}
\usepackage{amsmath,amsfonts,amsbsy,calrsfs}
\usepackage{pstricks,pst-node}
\usepackage[small,bf,hang]{caption2}
\usepackage{graphicx}

\usepackage{float}

\psset{unit=1.3cm,linewidth=.5pt,radius=.2}  

\def\hybrid{\topmargin -20pt    \oddsidemargin 0pt
        \headheight 0pt \headsep 0pt
        \textwidth 6.25in       
        \textheight 9.5in       
        \marginparwidth .875in
        \parskip 5pt plus 1pt   \jot = 1.5ex}

\hybrid

\newcommand{\beq}{\begin{equation}}
\newcommand{\eeq}{\end{equation}}
\newcommand{\bi}{\begin{itemize}}
\newcommand{\ei}{\end{itemize}}
\newcommand{\bt}{\begin{tabular}}
\newcommand{\et}{\end{tabular}}
\newcommand{\bc}{\begin{center}}
\newcommand{\ec}{\end{center}}

\def\theequation{\arabic{section}.\arabic{equation}}

\newcommand{\vev}[1]{\langle#1\rangle}

\newcommand{\be}{\begin{equation}}
\newcommand{\ee}{\end{equation}}
\newcommand{\bea}{\begin{eqnarray}}
\newcommand{\eea}{\end{eqnarray}}
\newcommand{\ba}{\begin{array}}
\newcommand{\ea}{\end{array}}

\def\bbox{{\,\lower0.9pt\vbox{\hrule \hbox{\vrule height 0.2 cm
\hskip 0.2 cm \vrule height 0.2 cm}\hrule}\,}}
\newcommand{\dsl}{\pa \kern-0.5em /}

\begin{document}

\begin{titlepage}
\begin{center}

\hfill MIT-CTP-4113

\vskip 2cm

{\Large \bf  A boundary stress tensor for higher-derivative\\[1.5ex] gravity
in AdS and Lifshitz backgrounds}

\vskip 1.5cm

{\bf Olaf Hohm and Erik Tonni} \\

\vskip 30pt

{\em \hskip -.1truecm Center for Theoretical Physics,\\ Massachusetts Institute of Technology,\\
Cambridge, MA 02139, USA \vskip 5pt }

{email: {\tt ohohm@mit.edu}, {\tt tonni@mit.edu}} \\

\end{center}

\vskip 1cm

\begin{center} {\bf ABSTRACT}\\[3ex]

\begin{minipage}{13cm}
\small
We investigate the Brown-York stress tensor for curvature-squared theories.
This requires a generalized Gibbons-Hawking term in order to establish a well-posed variational
principle, which is achieved in a universal way by reducing the number of derivatives through the introduction of an auxiliary tensor
field. We examine the boundary stress tensor thus defined for the special case of `massive gravity' in three dimensions, which augments the
Einstein-Hilbert term by a particular curvature-squared term. It is shown that one obtains finite results for physical parameters on AdS upon adding a `boundary cosmological constant' as a counterterm, which vanishes at the so-called chiral point. We derive known and new results, like the value of the central charges or the mass of black hole solutions, thereby confirming our prescription for the computation of the stress tensor. Finally, we inspect recently constructed Lifshitz vacua and a new black hole solution that is asymptotically Lifshitz, and we propose a novel and covariant counterterm for this case.

\end{minipage}

\end{center}


\vfill

January 2010

\end{titlepage}

\tableofcontents

\section{Introduction}\setcounter{equation}{0}
The AdS/CFT correspondence continues to provide novel perspectives on a wide range of physical problems. Here we just mention two developments, which are of relevance for the present paper. First, there have been interesting proposals to give a non-perturbative definition of quantum gravity in three dimensions via the dual CFT \cite{Witten:2007kt,Li:2008dq}. Second, it has become clear that variants of the AdS/CFT correspondence might provide important insights into strongly coupled field theories that are of direct physical relevance, as for instance in condensed matter systems or hydrodynamics. (For reviews see \cite{Hartnoll:2009sz,Herzog:2009xv,Rangamani:2009xk}.)

Concerning the latter development, one important new aspect is that
the relevant field theories are often non-relativistic, thus requiring a generalization of the usual AdS/CFT prescription. In particular, the background geometry of the gravitational dual has to deviate from the maximally symmetric Anti-de Sitter geometry, AdS$_{d+1}$, for the latter admits an isometry group that is directly related to the $d$-dimensional Lorentz invariance on the field theory side. One class of background geometries that are of interest in this respect are the so-called Lifshitz vacua which give rise to a different scaling of space and time
\cite{Kachru:2008yh}; see also \cite{Koroteev:2007yp}. (Another class are the Schr\"odinger space-times which feature in addition a Galilei symmetry \cite{Son:2008ye,Balasubramanian:2008dm}.) Unfortunately, these geometries do not appear as solutions of ordinary supergravity theories. This provides one motivation for considering higher-derivative theories of gravity, because these do permit Lifshitz vacua. (For applications of higher-derivative gravity in this context see, e.g., \cite{Adams:2008zk,Ghodsi:2009hg,Pang:2009ky}.) The first step towards AdS/CFT applications of higher-derivative theories consists of the computation of the boundary stress tensor \cite{Balasubramanian:1999re}. One aim of this paper is to compute the stress tensor for higher-derivative theories that are quadratic in the curvature.\footnote{For the stress tensor of gravitational theories that are dual to non-relativistic theories by virtue of appropriate matter couplings see \cite{Ross:2009ar}.}

Another motivation to compute the stress tensor for higher-derivative gravity is the special case of the so-called new massive gravity (NMG) 
\cite{Bergshoeff:2009hq,Bergshoeff:2009aq}, see also 
\cite{Andringa:2009yc,Bergshoeff:2009tb,Bergshoeff:2009fj}.
Specifically, this is a particular curvature-squared theory in three dimensions, which has been discussed as an interesting candidate for a consistent theory of quantum gravity. The knowledge of the stress tensor allows one to compute the mass or energy of gravitational solutions, in particular of black holes. This is of importance because it has been clear since the attempts of \cite{Li:2008dq} that there are potential conflicts in theories of this type with having unitary graviton (bulk) modes and positive mass for the so-called BTZ black holes. There are more black hole solutions in NMG, and so we compute here also their masses by use of the stress tensor.
An independent reason to consider NMG is the recent observation that it permits Lifshitz vacua and black holes \cite{AyonBeato:2009nh}. Thus, here we have the opportunity to address the question of how to compute mass and other physical parameters in such backgrounds.

Before we proceed, let us briefly recall the general definition of the Brown-York stress tensor \cite{Brown:1992br,Brown:2000dz}. Suppose one starts from a gravitational action $S$ that leads to a well-posed variational principle in the following sense: in order to derive the required bulk equations of motion it is sufficient to set the variation of the fields at the boundary equal to zero, in particular the variation of the boundary metric, $\delta\gamma_{ij}=0$, but not their normal derivatives. Then one can define the boundary stress tensor as the variation with respect to the boundary metric,
 \bea\label{stressten0}
  T^{ij} \ = \ \frac{2}{\sqrt{-\gamma}}\frac{\delta S}{\delta \gamma_{ij}}\;.
 \eea
This gives rise to conserved charges and, in particular, to a definition of the mass or energy as a $(d-1)$--dimensional integral over $T_{00}$.
Unfortunately, gravitational theories as usually formulated do not permit a well-posed variational principle, and so in this case (\ref{stressten0}) is not even meaningful. For the second-order Einstein-Hilbert theory this can be remedied by adding the Gibbons-Hawking term as a boundary action. However, due to their higher-derivative nature, for generic theories involving quadratic or higher powers of the curvature there is no corresponding boundary term that leads to a well-posed variational principle \cite{Madsen:1989rz}, even though in special cases this might be different (for some exceptions see \cite{Myers:1987yn,Grumiller:2008ie}). The various aspects of this problem have been discussed extensively in the literature for the special class of $f(R)$ theories, which involve only the scalar curvature, see \cite{Dyer:2008hb} for a recent survey. Here, the problem can be circumvented by introducing an auxiliary scalar $\phi$, which allows one to reformulate the action as a second-order `dilaton-like' gravity with a $f(\phi) R$ term. Due to this effective reduction of the number of derivatives, a generalized Gibbons-Hawking term does exist and is of the form $f(\phi) K$, where $K$ is the trace of the extrinsic curvature. Because the variation of $\phi$ is then also taken to be zero at the boundary, this leads to a well-posed variational principle.
Equivalently, one may establish a well-defined variational principle by imposing the additional boundary condition that the on-shell value of 
$\phi$, i.e., the scalar curvature $R$, is not varied. 

It is fairly obvious that this strategy can be generalized to generic higher-derivative theories of gravity, upon introducing more auxiliary fields that have a non-trivial tensor structure, see for instance \cite{Balcerzak:2008bg}. However, a systematic treatment of the Brown-York stress tensor (\ref{stressten0}) computed in this way is so far lacking. (For related approaches see 
\cite{Nojiri:1999nd,Cvetic:2001bk,Astefanesei:2008wz,Cremonini:2009ih}.) 
Here, we are going to reformulate generic curvature-squared theories in such a way that there is a generalized Gibbons-Hawking term so that (\ref{stressten0}) can be applied. However, strictly speaking this procedure is not unique, because there are different choices for the auxiliary fields and it is not clear a priori that these different choices lead to physically equivalent answers. We propose a specific set of auxiliary fields that is in some sense minimal, and we verify that it leads to physically sensible answers for NMG by comparing with quantities that have been computed by other methods.

This paper is organized as follows. In sec.~2 we first determine the generalized Gibbons-Hawking term for NMG and then extend these results to arbitrary curvature-squared theories. In sec.~3 we compute the boundary stress tensor and determine the counterterm that leads to finite results for asymptotically AdS$_3$ solutions. In the following sections we apply this renormalized stress tensor to NMG. We compute in sec.~4 the central charges, the mass of BTZ black holes and of new black holes at the `chiral point'. In sec.~5 we turn to the Lifshitz vacua and an asymptotically Lifshitz black hole solution, and we show that by virtue of the auxiliary fields a novel covariant counterterm can be constructed. We conclude in sec.~6. Some technical details about the ADM decomposition and auxiliary fields in arbitrary dimensions can be found in the appendices.

\section{The generalized Gibbons-Hawking term}
In this section we are going to formulate general curvature-squared theories in such a way that they allow a well-posed variational principle. As outlined in the introduction, this requires an auxiliary field in order to cast the action into a form which is second-order in derivatives. Before turning to the general theory in arbitrary dimensions, we first illustrate the method with the special case of massive gravity in three dimensions as this is the case we will analyze in more detail later.

\subsection{Massive gravity in $D=3$}
The massive $D=3$ gravity theory constructed in \cite{Bergshoeff:2009hq} is defined by the Lagrangian
 \bea\label{CNMG}
  {\cal L}_{\rm NMG} \ = \ \frac{1}{\kappa^2}\sqrt{-g}\left[\sigma R+\frac{1}{m^2}\left(R^{\mu\nu}
  R_{\mu \nu}-\frac{3}{8}R^2\right)-2\lambda m^2\right]\;,
 \eea
where $\sigma=\pm 1$ controls the sign of the Einstein-Hilbert term, $m$ is a mass parameter and $\lambda$ a cosmological parameter. 
One remarkable feature of this theory is that, despite its higher derivatives, it propagates massive graviton modes unitarily about some of its maximally symmetric vacua for $\sigma=-1$. Given our conventions (which are those of \cite{Bergshoeff:2009aq}), this means that the sign of the Einstein-Hilbert term is the opposite of the sign in higher dimensions. In the following we will also refer to the choices $\sigma=+1$ and $\sigma=-1$ as the `right-sign' and `wrong-sign' Einstein-Hilbert term, respectively. One question we will address below is whether the requirement that black hole solutions have positive mass leads to the same sign of the Einstein-Hilbert term as the requirement that the graviton modes be unitary (i.e.~have positive energy).

In the fourth-order form there is no well-posed variational principle for which only the variations $\delta g_{\mu\nu}$ of the metric but not their (normal) derivatives are set to zero at the boundary. To remedy this we cast the action into second-order form by introducing
an auxiliary field $f_{\mu\nu}$,
 \bea\label{1CNMG}
  {\cal
  L}\ = \ \frac{1}{\kappa^2}\sqrt{-g}\left[\sigma R+f^{\mu\nu}G_{\mu\nu}-\frac{1}{4}m^2
  \left(f^{\mu \nu}f_{\mu\nu}-f^2\right)-2\lambda m^2\right]\;,
 \eea
where $f=g^{\mu\nu}f_{\mu\nu}$ and
 \bea
  G_{\mu\nu} \ = \  R_{\mu\nu}-\frac{1}{2}R\, g_{\mu\nu}
 \eea
is the Einstein tensor. As far as the bulk dynamics is concerned, this action is equivalent to (\ref{CNMG}), for the algebraic field equations of $f_{\mu\nu}$,
 \bea\label{fonshell}
  f_{\mu\nu} \ = \ \frac{2}{m^2}\left(R_{\mu\nu}-\frac{1}{4}R\,
  g_{\mu\nu}\right)\;,
 \eea
allow one to eliminate this auxiliary field, which gives back (\ref{CNMG}). This choice of auxiliary field proved also to be useful in order to analyze the physical content of NMG \cite{Bergshoeff:2009hq,Bergshoeff:2009aq}. 

Let us now inspect the equations of motion and the variational principle.
The `Einstein equations' obtained by varying the metric read
 \bea\label{bulkeinstein}
 \begin{split}
  &\sigma G_{\mu\nu}+\lambda m^2 g_{\mu\nu}
  -\frac{1}{2}m^2\Big[f_{\mu}{}^{\rho}f_{\nu\rho}-ff_{\mu\nu}-\frac{1}{4}g_{\mu\nu}
  \big(f^{\rho\sigma}f_{\rho\sigma}-f^2\big)\Big] \\
  &+2f_{(\mu}{}^{\rho}G_{\nu)\rho}+\frac{1}{2}Rf_{\mu\nu}
  -\frac{1}{2}fR_{\mu\nu}
  -\frac{1}{2}g_{\mu\nu}\,f^{\rho\sigma}G_{\rho\sigma}\\
  &+\frac{1}{2}\Big[D^2f_{\mu\nu}-2D^{\rho}D_{(\mu}f_{\nu)\rho}+D_{\mu}D_{\nu}f
  +(D^{\rho}D^{\sigma}f_{\rho\sigma}-D^2 f)g_{\mu\nu}\Big] \ = \ 0\;.
 \end{split}
 \eea
Inserting (\ref{fonshell}) into this equation one obtains the 4th order Einstein equations
as given in \cite{Bergshoeff:2009hq}. It is this equivalent form which we have used for the computations to be discussed below in which we confirmed solutions of the NMG field equations.
We should note that the explicit expression (\ref{bulkeinstein}) involves an ambiguity, for
it depends on the question which index structure of $f$ we consider as fundamental and which we regard as obtained by raising and lowering indices with $g_{\mu\nu}$. For definiteness,  here we have chosen $f_{\mu\nu}$ as fundamental field. One may check that the fourth-order equations are independent of this choice. However, the boundary stress tensor does depend on such a choice, and we will come back to this point below.

If we vary now the action about a solution of the bulk equations of motion (\ref{bulkeinstein}),
the terms linear in the variations naturally vanish up to boundary terms. To be more precise, the variation reads
 \bea\label{boundvar}
  \delta S_{\rm NMG} \ = \ \frac{1}{\kappa^2}\int_{M_3} d^3 x \sqrt{-g}\,D_{\mu}v^{\mu}\;,
 \eea
where
 \bea\label{v}\nonumber
  v^{\mu} &=& \sigma w^{\mu}-\frac{1}{2}f w^{\mu}+
  f^{\rho\sigma}g^{\mu\nu}D_{\rho}\left(\delta g_{\sigma\nu}\right)
  -\frac{1}{2}f^{\rho\sigma}D^{\mu}\left(\delta g_{\rho\sigma}\right)
  -\frac{1}{2}f^{\mu\nu}g^{\rho\sigma}D_{\nu}\left(\delta g_{\rho\sigma}\right)\\ 
  &&+\frac{1}{2}\Big[D^{\mu}f^{\nu\rho}-2D^{\nu}f^{\mu\rho}
  +g^{\mu\nu}D^{\rho}f+g^{\nu\rho} D_{\sigma}f^{\sigma\mu} 
  -g^{\nu\rho} D^{\mu}f\Big]
  \delta  g_{\nu\rho}\;,
 \eea
and we introduced 
\bea
  w^{\mu} \ = \ g^{\mu\nu}D^{\rho}(\delta g_{\nu\rho})-g^{\rho\sigma}D^{\mu}
  (\delta g_{\rho\sigma})\;,
\eea
which defines the total derivative appearing in the variation of the pure Einstein-Hilbert term. 
In order to have a well-defined variational principle, all terms with a bare $\delta g_{\mu\nu}$ may be set to zero at the boundary, so these terms in (\ref{v}) can be ignored for the moment (though they will contribute later to the boundary stress tensor). To get rid of the remaining terms, we have to add local boundary terms, to which we turn now.

In order to define the boundary terms generalizing the Gibbons-Hawking term, we choose coordinates $x^{\mu}=(x^{i},r)$,
corresponding to a slicing of the bulk $M_3$ by two-dimensional Lorentzian
submanifolds $M_r$ for each value of the radial coordinate $r$, each
slice being parameterized by the coordinates $x^{i}$. One should
think of $M_r$ in the limit $r\rightarrow\infty$ as the
two-dimensional boundary, which is the space on which the dual CFT will be defined. Correspondingly, we make an ADM-like split of the metric,
 \bea\label{ADMelement}
  ds^2 \ = \ N^2dr^2+\gamma_{ij}\left(dx^{i}+N^i dr\right)\left(dx^j+N^j dr\right) \;,
 \eea
where $\gamma_{ij}$ defines the boundary metric, while $N$ and $N_{i}$ denote lapse and shift function, respectively. (For more details on the ADM decomposition we refer to appendix A.) Using Stokes' theorem and
$\sqrt{-g}=\sqrt{-\gamma}N$, (\ref{boundvar}) can also be written as
 \bea\label{covvar}
  \delta S_{\rm NMG} \ = \ \frac{1}{\kappa^2}\int_{\partial M_3} d^2
  x\sqrt{-\gamma}\;n_{\mu}v^{\mu}\;, \qquad
  n_{\mu} \ = \ (0,0,N)\;,
 \eea
where $n_{\mu}$ is the normal vector on the hypersurfaces $M_r$. The three-dimensional geometry is fully encoded in the \textit{intrinsic} boundary geometry determined by the boundary metric together with the \textit{extrinsic} curvature tensor defined in terms of the normal vectors,
 \begin{eqnarray}
  K_{\mu\nu} &=&
  -\frac{1}{2}(D_{\mu}n_{\nu}+D_{\nu}n_{\mu})\;.
 \end{eqnarray}
The relevant `boundary part' then reads
 \bea\label{extrinsic}
  K_{ij} \ = \ \Gamma_{ij}^rN \ = \
  -\frac{1}{2N}(\partial_r\gamma_{ij}-\nabla_{i}N_{j}-\nabla_{j}N_i)\;,
 \eea
while the other components, $K_{rr}$ and $K_{ir}$, can be expressed in terms of $K_{ij}$, i.e., the latter tensor encodes the full information about the extrinsic geometry.
Here and in the following $\nabla_{i}$ denotes the covariant derivative with respect to the Levi-Civita connection determined by the boundary metric $\gamma_{ij}$.
So far we have not fixed any gauge, since the lapse and shift functions in (\ref{ADMelement}) are completely arbitrary. For the computation it turns out to be convenient to perform a gauge-fixing according to
 \bea\label{gaugefix}
  N_i \ = \ 0\;, \qquad N \ = \ 1 \;,
 \eea
corresponding to Gaussian normal coordinates, for which $K_{ij}=-\frac{1}{2}\partial_r\gamma_{ij}$. In order to reconstruct the completely general result at the end of the computation, i.e., to undo the gauge-fixing, we simply have to replace $\partial_r\gamma_{ij}$ by $-2K_{ij}$ according to (\ref{extrinsic}), as we review  in the appendix.

Next we have to express the radial component $v^r$ of (\ref{v}), which is the relevant one for the boundary part of the variation, in terms of $K_{ij}$ and the intrinsic boundary geometry. For this we have to choose a $2+1$ split also for the auxiliary $f^{\mu\nu}$,
 \bea\label{fdeco}
  f^{\mu\nu} \ = \ \left(\begin{array}{cc} f^{ij} &
  h^i \\ h^j & s \end{array}\right)\;.
 \eea
Here we have to keep in mind that even after the gauge-fixing (\ref{gaugefix}) the components $s$ and $h^{i}$ are generally non-zero.
A straightforward computation then shows that the variation (\ref{boundvar}) can be written as
 \bea\label{fullvar}
  \delta S_{\rm NMG} &=& \frac{1}{\kappa^2}\int_{\partial M_3} d^2 x\sqrt{-\gamma}\, 
  v^{r}  \\ \nonumber
  &=&\frac{1}{\kappa^2}\int_{\partial M_3} d^2 x\sqrt{-\gamma}\left[ 2\sigma \gamma^{ij} 
  \delta K_{ij}
  +f^{ij}\delta K_{ij}
  -f\gamma^{ij}\delta K_{ij}\right]\;,
 \eea
where here and in the following we denote by slight abuse of notation $f=\gamma^{ij}f_{ij}$, in contrast to the covariant trace above.
We ignored all terms with a bare $\delta\gamma_{ij}$. Moreover, we have omitted a term proportional to $h^{k}\gamma^{ij}\nabla_{k}(\delta\gamma_{ij})$, since it contains only a boundary derivative of $\delta\gamma_{ij}$, and so upon partial integration (using that the boundary of the boundary is empty) it may be brought to a form with a bare $\delta\gamma_{ij}$. From
(\ref{fullvar}) it is now straightforward to read off the generalized Gibbons-Hawking term,
 \bea\label{gGH}
  S_{\rm GGH} \ = \ \frac{1}{\kappa^2}\int_{\partial M_3} d^2 x\sqrt{-\gamma}\left(
  -2\sigma K-\hat{f}^{ij}K_{ij}+\hat{f}K\right)\;,
 \eea
where $K=\gamma^{ij}K_{ij}$ is the trace of the extrinsic curvature.
Moreover, here we use the following combinations of the components of $f^{\mu\nu}$,
 \bea\label{hatted}\nonumber
  \hat{f}^{ij} &=& f^{ij} +2h^{(i}N^{j)}+sN^{i}N^{j}\;, \\
  \hat{h}^i &=& N(h^{i}+sN^{i})\;, \\  \nonumber
  \hat{s} &=& N^2s\;,
 \eea
which are fixed by the requirement that they transform covariantly and reduce to the fields  defined in (\ref{fdeco}) for the gauge-fixing (\ref{gaugefix}) to normal coordinates. (See appendix A for the derivation.) The term proportional to $\sigma$ is the well-known Gibbons-Hawking term for pure gravity, to which the boundary action reduces in the limit $f^{\mu\nu}\equiv 0$ corresponding to $m^2\rightarrow\infty$.
Let us point out again that after the addition of (\ref{gGH}) we have a well-posed variational principle by virtue of the fact that the variation of $f^{\mu\nu}$ (as for any other `bare' field) can be taken to be zero at the boundary, $\delta f^{\mu\nu}=0$. This is in contrast to the original fourth-order formulation without further boundary conditions. In that case the components of $f^{\mu\nu}$ entering (\ref{gGH}) would be replaced by their on-shell values determined by the full bulk curvature tensor, whose variation in turn gives rise to additional contributions involving $K_{ij}$, thus annihilating the original cancellation of these terms. In other words, in order to establish a well-posed variational principle in the fourth-order formulation one would have to impose the boundary condition that  precisely those combinations of the bulk curvature tensor that determine 
$f^{\mu\nu}$ on-shell are kept fixed during the variation. It should be clear from this discussion 
that the role of the auxiliary field is to provide a technical simplification in that it allows for a clear separation between the objects which have to be varied and those which have to be kept fixed.

\subsection{Curvature-squared theories in general dimensions}
Let us now see whether the results of the previous subsection generalize to curvature-squared theories in arbitrary dimensions. The main
difference to the three-dimensional case is that in higher dimensions there can also be a term involving the Riemann tensor. Thus we consider Lagrangians of the form  
\begin{equation}
\label{lagrangian riemann squared}
\mathcal{L}\,=\, \sqrt{-g}\left(
a_1 R^{\mu\nu\rho\sigma} R_{\mu\nu\rho\sigma} +a_2 R^{\mu\nu} R_{\mu\nu}
+a_3 R^2\right)\;.
\end{equation}
Without the term quadratic in the Riemann tensor a formulation second-order in derivatives could be achieved in complete analogy to (\ref{1CNMG}), upon introducing an auxiliary  symmetric 2-tensor, for which the free coefficients are encoded in the coefficients of
the terms quadratic in the auxiliary fields. However, the presence of the full Riemann tensor requires a new 4-index auxiliary field.
As we pointed out before, there is an ambiguity of which set of auxiliary fields to choose. For instance, one could simply add an auxiliary field
$\phi^{\mu\nu\rho\sigma}$, carrying the symmetries of the Riemann tensor, to the symmetric 2-tensor. However, this choice
is non-minimal in that the 4-index field by itself is sufficient to cast the action into a form which is second-order in derivatives. In fact, the
Lagrangian
 \bea\label{gensquare}
  {\cal L} \ = \ \sqrt{-g}\left(\phi^{\mu\nu\rho\sigma}R_{\mu\nu\rho\sigma}+b_1\phi^{\mu\nu\rho\sigma}\phi_{\mu\nu\rho\sigma}
  +b_2 \phi^{\mu\nu}\phi_{\mu\nu}+b_3 \phi^2\right)\;,
 \eea
with $\phi_{\mu\nu}$ and $\phi$ denoting the single and double trace,
leads to generic curvature-squared actions of the form (\ref{lagrangian riemann squared})  upon integrating out $\phi_{\mu\nu\rho\sigma}$, with the three free coefficients determined by the coefficients
$b_1$, $b_2$ and $b_3$ (see appendix B for more details). 
In three dimensions $\phi_{\mu\nu\rho\sigma}$ carries the same number of components as $f_{\mu\nu}$
--- for the same reasons that the Riemann tensor is equivalent to the Ricci tensor ---, and so in this case (\ref{gensquare}) is equivalent to 
our previous ansatz whose validity we confirm below.

Next we investigate the variational principle of (\ref{gensquare}). The on-shell variation contains a boundary term originating from
 \bea
  \delta S\Big|_{\rm on-shell} \ = \ \int d^{d+1} x\,\sqrt{-g}\,\phi^{\mu\nu}{}_{\rho}{}^{\sigma}\delta R_{\mu\nu}{}^{\rho}{}_{\sigma}
  \ = \ \int d^{d+1} x\,\sqrt{-g}\, D_{\mu}v^{\mu}\;,
 \eea
with
 \bea
   v^{\mu} \ = \ -2\phi^{\mu\nu\rho\sigma}D_{\rho}(\delta g_{\nu\sigma})
   +2D_{\rho}\phi^{\mu\nu\rho\sigma}\delta g_{\nu\sigma}\;.
 \eea
Here we ignored terms that are zero on-shell, while the second term with a bare 
$\delta g_{\mu\nu}$ is irrelevant for the variational principle.
Next we rewrite this expression in a ADM decomposition adapted to the boundary, using a
gauge-fixing to normal coordinates. We obtain for the radial component 
 \bea
  v^{r} \ = \ 4\phi^{ij}\delta K_{ij}\;,
 \eea
where $\phi^{ij}=\phi^{irjr}$. Thus the Gibbons-Hawking term takes the following form,
 \bea\label{gengenGH}
   S_{\rm GGH} \ = \ -4\int_{\partial M_{d+1}}d^{d}x\,\sqrt{-\gamma}\,\hat{\phi}^{ij}K_{ij}\;,
 \eea
with the covariant version of the auxiliary field given by $\hat{\phi}^{ij}=N^2\phi^{irjr}$, see appendix A.
Remarkably, this form is completely universal in that it is independent of the details of the theory, i.e., on the precise coefficients of the curvature-squared terms and the space-time dimension.

As a consistency check we next verify that upon specializing (\ref{gengenGH}) to $D=3$, we obtain the result of the previous subsection.
In $D=3$ the Einstein tensor is related to the full Riemann tensor via
 \bea
  R_{\mu\nu\rho\sigma} \ = \ \varepsilon_{\mu\nu\lambda}\varepsilon_{\rho\sigma\tau}G^{\lambda\tau}\;.
 \eea
Therefore, the relevant term in the $D=3$ first-order action, $f^{\mu\nu}G_{\mu\nu}$, reduces to the one in (\ref{gensquare}) by virtue of the identification
$f^{\mu\nu}=\varepsilon^{\mu\rho\sigma}\varepsilon^{\nu\lambda\tau}\phi_{\rho\sigma\lambda\tau}$. The boundary part is then given by
 \bea\label{fphi}
  \hat{f}^{ij} \ = \ 4\big(\hat{\phi}^{ij}-\gamma^{ij}\hat{\phi}\big)\;,
 \eea
where $\hat{\phi}$ denotes the trace with respect to $\gamma_{ij}$. Inserting (\ref{fphi}) into (\ref{gGH}) we obtain precisely (\ref{gengenGH}). Thus the two forms of the generalized Gibbons-Hawking term in $D=3$ are equivalent.

\section{The boundary stress tensor and its counterterms}
So far we have shown that generic curvature-squared theories in arbitrary dimensions 
admit a reformulation that allows for a generalized Gibbons-Hawking term which leads to a 
well-posed variational principle. The boundary stress tensor is then well-defined and determined according to (\ref{stressten0}). Here we are first giving this boundary stress tensor for NMG and then for the generic theory in higher dimensions. Finally, we extend the result for NMG by a counterterm, which is required in order to obtain finite results.\footnote{For alternative methods to compute conserved charges in NMG we refer to \cite{Clement:2009gq,Oliva:2009ip}.}     

Taking the full boundary variation resulting from (\ref{v}) and the generalized Gibbons-Hawking term (\ref{gGH}) into account, we find
 \bea\label{stressten}\nonumber
  8\pi G\, T^{ij} &=& \sigma \left(K^{ij} - K\gamma^{ij}\right)-\frac{1}{2}\hat{f}K^{ij}
  -\nabla^{(i}   \hat{h}^{j)}
  +\frac{1}{2}{\cal D}_r
  \hat{f}^{ij}
  -K^{(i}{}_{k}\hat{f}^{j)k}
  +\frac{1}{2}\hat{s}K^{ij}\\
  &&+\gamma^{ij}\left(\nabla_{k}\hat{h}^k-\frac{1}{2}\hat{s}K
  +\frac{1}{2}\hat{f}K-\frac{1}{2}{\cal D}_r \hat{f}\right)\;,
 \eea
where we introduced Newton's constant via $\kappa^2=16\pi G$.
Here we have used again that the computation is most easily done in normal coordinates for which (\ref{gaugefix}) holds. After this, the result can be `covariantized' by introducing the full extrinsic curvature $K_{ij}$ and replacing the components of $f^{\mu\nu}$ by the hatted quantities (\ref{hatted}) and introducing the `covariant $r$-derivative'
${\cal D}_r$. (For more details we refer to appendix A.) As before, we observe that the terms proportional to $\sigma$ correspond to those of the standard boundary stress energy tensor for pure gravity \cite{Brown:1992br}, and the full
$T^{ij}$ reduces to this one in the limit $m^2\rightarrow\infty$, for which $f^{\mu\nu}=0$.

We have to comment on the following subtlety of the computation. There is an ambiguity related to the definition of the auxiliary field $f$, namely whether we regard the fundamental field as a tensor with lower, upper or mixed indices. While the bulk equations of motion are independent of this choice, as it should be, the boundary stress tensor (\ref{stressten}) does depend on it, for it is this choice that determines how the generalized Gibbons-Hawking term depends on the boundary metric $\gamma_{ij}$. For instance, if we regard the fundamental field as an object with lower indices, the relevant part of the Gibbons-Hawking term reads
 \bea\label{GHexplicit}
  \int d^2x\sqrt{-\gamma}\,f^{ij}K_{ij} \ \equiv \ 
  \int d^2x\sqrt{-\gamma}\,\gamma^{ik}\gamma^{jl}f_{ij}K_{kl}\;,
 \eea
and so the variational derivative $\delta S/\delta\gamma^{ij}$ will contain a term proportional to $f_{i}{}^{k}K_{jk}$. On the other hand, if we regard $f^{ij}$ with upper indices as fundamental, the term in (\ref{GHexplicit}) does not require a metric to raise indices and therefore the variational derivative does not lead to the same term. Thus, even though both forms of the Gibbons-Hawking term are invariant under boundary-preserving diffeomorphisms, they lead to different contributions to the stress tensor.      
There is no way to fix this ambiguity a priori, but one may use the following heuristic reasoning. In order to compute the one-point function $\vev{T_{ij}}$ of the energy-momentum tensor in the AdS/CFT correspondence, one has to inspect its coupling to a source (namely the graviton modes) according to $h^{ij}T_{ij}$. Now, these modes are exponentially growing or decaying as functions of the radial coordinate unless its index structure is $h_{i}{}^{j}$, in which case it is finite 
\cite{Liu:1998bu}.  
This suggests that also the fundamental form of the auxiliary field is 
$f_{\mu}{}^{\nu}$.\footnote{We are grateful to Daniel Grumiller for suggesting this argument to us.} 
It is this choice that has been used in the computation of (\ref{stressten}), and below we will show that it leads to the correct value of the central charges, that have been determined before by other methods.

For completeness we record here also the stress tensor in generic dimensions, which is  derived from the action (\ref{gensquare}) together with the Gibbons-Hawking term (\ref{gengenGH}) by use of (\ref{stressten0}). It reads
 \bea\label{stresshigher}
  T^{ij} \ = \ 4\left(2\nabla_{k}\hat{\phi}^{ikj}+{\cal D}_{r}\hat{\phi}^{ij}
  +K_{kl}\hat{\phi}^{ikjl}-K\hat{\phi}^{ij}-\gamma^{ij}\,\hat{\phi}^{kl}K_{kl}\right)\;,
 \eea
where we have not included the contribution of the Einstein-Hilbert term or other possible couplings. 
(We refer to appendix A for the hatted quantities.) Let us stress that, as it stands, 
(\ref{stresshigher}) is most likely incomplete because generically one still needs to add a counterterm in order to obtain finite results. Here we are not going to further investigate the stress tensor in arbitrary dimensions, but focus instead on (\ref{stressten}) for NMG.     

Next we evaluate the boundary stress tensor of NMG for the maximally symmetric AdS background in order to determine the counterterm that renders the result for physical parameters finite. In general, following Brown and York \cite{Brown:1992br}, the knowledge of the stress tensor allows one to construct the conserved charges associated to the Killing vectors $\xi^i$ of the boundary geometry according to \cite{Balasubramanian:1999re}
 \bea\label{charges}
  Q_{\xi} \ = \ \int_{\Sigma}dx\sqrt{\rho}\,u^{i}\,T_{ij}\,\xi^{j}\;.
 \eea
The functions appearing here refer to an ADM-like split, but now of the two-dimensional boundary metric,
 \bea
  \gamma_{ij}dx^{i}dx^{j} \ = \ -N_{\Sigma}^2dt^2+\rho\left(dx+N_{\Sigma}^{*}dt\right)^2 \;,
 \eea
where $\Sigma$ denotes a space-like `surface' with one-dimensional metric $\rho$ and time-like normal vector $u^{i}$.
The charges (\ref{charges}) allow us to define mass and momentum as boundary integrals,
 \bea\label{massform}
  M \ = \ \int_{\Sigma}dx\sqrt{\rho}\,N_{\Sigma}\,T_{ij}\,u^{i}\,u^{j}\;, \qquad
  P_{x} \ = \ \int_{\Sigma}dx\sqrt{\rho}\,\rho\, u_{i}\,T^{xi}\;.
 \eea

Let us now give these formulas explicitly for AdS$_3$, which we parameterize with standard coordinates,
 \bea\label{ads3}
  ds^2 \ = \ \frac{\ell^2}{r^2}dr^2+\frac{r^2}{\ell^2}\left(-dt^2+dx^2\right)\;,
 \eea
which implies that the shift functions are zero and the lapse is $N(r)=\ell/r$.
The auxiliary field is determined by its field equation in terms of the AdS metric, which implies in particular
 \bea\label{fads3}
  \hat{f}^{ij}=-\frac{1}{m^2\ell^2}\gamma^{ij}\;, \qquad
  \hat{h}^i \ = \ 0\;, \qquad
  \hat{s} \ = \ -\frac{1}{m^2\ell^2}\;.
 \eea
In the coordinates (\ref{ads3}), the mass or energy of an asymptotically AdS solution and its angular momentum, respectively, are given by
 \bea\label{massmomentum}
  M \ = \ \int dx\, T_{00}\;, \qquad
  J \ = \ -\int dx\, T_{10}\;.
 \eea

We next have to evaluate $T_{00}$ by inserting (\ref{ads3}) and (\ref{fads3}) into
(\ref{stressten}). One finds 
 \bea\label{T00}
  8\pi G\,T_{00} \ = \ -\frac{r^2}{\ell^3}\left(\sigma+\frac{1}{2m^2\ell^2}\right)\;,
 \eea
which diverges once we go to the boundary by sending $r\rightarrow \infty$. This can be remedied by adding a local pure boundary term as a counterterm to the action. The leading  term in a derivative expansion involving only the boundary metric is a `cosmological constant' term, and it turns out that this is sufficient also for the present theory. Specifically, the new action
 \bea
  S_{\rm total} \ = \ S_{\rm NMG} + S_{\rm GGH}
  -\frac{1}{8\pi G\ell}\left(\sigma+\frac{1}{2m^2\ell^2}\right)\int d^2 x\;\sqrt{-\gamma}
 \eea
gives rise to a renormalized stress tensor
 \bea\label{regten}
  T_{ij}^{\rm ren} \ = \ T_{ij}-\frac{1}{8\pi G\ell}\left(\sigma +\frac{1}{2m^2\ell^2}\right)
  \gamma_{ij}\;.
 \eea
By construction, the extra contribution cancels the divergent term in (\ref{T00}), leaving vanishing mass, $M=0$, for the AdS groundstate, as required.

\section{Dual CFT and black hole parameters}
After having determined the boundary stress tensor and the local counterterm rendering physical parameters finite, we are now in a position to compute various interesting quantities. We begin with the central charges of the hypothetical CFT at the boundary,
applying the method of \cite{Balasubramanian:1999re}, and then turn to the computation of mass and angular momentum for various black hole solutions.

\subsection{Central charges}
The starting point for this computation is the fact that the central charges encode the anomalous transformation behaviour of the energy-momentum tensor under conformal transformations, i.e., the Weyl anomaly \cite{Brown:1986nw,Henningson:1998gx,deHaro:2000xn,Imbimbo:1999bj}. 
Specifically, if we choose light-cone coordinates
$x^{\pm}$, the conformal transformations take the standard form
 \bea
  \delta x^{+} \ = \ -\xi^{+}(x^{+})\;, \qquad \delta x^{-} \ = \ -\xi^{-}(x^{-})\;,
 \eea
under which the stress tensor transforms as
 \bea\label{centraldef}
  \delta T_{++} \ = \ {\cal L}_{\xi}T_{++}-\frac{c}{24\pi}\partial_+ ^3\xi^{+}\;, \qquad
  \delta T_{--} \ = \ {\cal L}_{\xi}T_{--}-\frac{c}{24\pi}\partial_{-}^3\xi^{-}\;.
 \eea
Here, ${\cal L}_{\xi}$ denotes the Lie derivative, i.e., these terms represent the
covariant part of the transformation. Now, the stress tensor (\ref{stressten}) computed from the gravitational action does transform covariantly under the boundary-preserving diffeomorphisms discussed in the appendix.\footnote{In fact, the precise form of (\ref{stressten}) has been fixed by requiring a covariant transformation behaviour.} However, non-trivial central charges and thus anomalous transformation rules emerge in this picture
due to the fact that the asymptotic symmetry group of AdS$_3$ is larger than these restricted diffeomorphisms. The boundary conditions defining the asymptotic form of AdS$_3$ have been given by Brown and Henneaux \cite{Brown:1986nw} (though at the so-called chiral point these can be relaxed to logarithmic boundary conditions \cite{Liu:2009kc,Grumiller:2008es}, leading to the notion of `log gravity'). Starting from the AdS$_3$ metric with the boundary part written in light-cone coordinates,
 \bea
  ds^2 \ = \ \frac{\ell^2}{r^2}dr^2-r^2dx^{+}dx^{-}\;,
 \eea
the asymptotic form of the metric is given by
 \bea
  g_{+-} &=&  -\frac{r^2}{2}+{\cal O}(1)\;, \qquad
  g_{++} \ = \  {\cal O}(1)\;, \qquad\;
  g_{--} \ = \ {\cal O}(1)\;, \\ \nonumber
  g_{rr} &=& \frac{\ell^2}{r^2}+{\cal O}\left(1\over r^4\right)\;, \quad\;
  g_{+r} \ = \ {\cal O}\left(1\over r^3 \right)\;, \quad
  g_{-r} \ = \ {\cal O}\left(1\over r^3 \right)\;.
 \eea
The diffeomorphisms that leave this asymptotic form invariant are parametrized by the following vector fields
 \bea\label{BHdiff}\nonumber
  X^{+} &=&  \xi^{+}(x^+)+\frac{\ell^2}{2r^2}\partial_{-}^2\xi^{-}(x^{-})\;, \\ 
  X^{-} &=&  \xi^{-}(x^{-})+\frac{\ell^2}{2r^2}\partial_{+}^2\xi^{+}(x^{+})\;, \\ \nonumber
  X^r   &=& -\frac{r}{2}\left(\partial_+\xi^{+}+\partial_{-}\xi^{-}\right)\;,\nonumber
 \eea
as one may verify with (\ref{diff}). We observe that the $r$-component also depends on the
boundary coordinates $x^{i}$, and so these diffeomorphisms do not belong to the class of boundary-preserving diffeomorphisms under which the stress tensor transforms covariantly.

Under (\ref{BHdiff}) the AdS$_3$ metric transforms as
 \bea
  \delta g_{++} \ = \ -\frac{\ell^2}{2}\partial_{+}^3\xi^{+}\;, \qquad
  \delta g_{--} \ = \ -\frac{\ell^2}{2}\partial_{-}^3\xi^{-}\;,
 \eea
with all other components staying invariant. The extrinsic curvature $K_{ij}$ of AdS$_3$
and its trace stay invariant. The diagonal components of the auxiliary field $f^{ij}$ also transform non-trivially,
 \bea
   \delta f^{++} \ = \ -\frac{2}{m^2r^4}\partial_{-}^3\xi^{-}\;, \qquad
   \delta f^{--} \ = \ -\frac{2}{m^2r^4}\partial_{+}^3\xi^{+}\;.
 \eea
With these ingredients the transformation of, say, $T_{++}$ can be computed from
(\ref{regten}),
 \bea
  8\pi G\,\delta T_{++} \ = \  -\frac{\ell}{2}\partial_{+}^3\xi^{+}
  \left(\sigma+\frac{1}{2m^2\ell^2}\right)\;,
 \eea
which implies for the central charge by use of (\ref{centraldef})
 \bea\label{central}
   c \ = \ \frac{3\ell}{2 G} \left(\sigma+\frac{1}{2m^2\ell^2}\right)\;.
 \eea
This coincides with the value computed in \cite{Bergshoeff:2009aq,Liu:2009bk} by other methods (based on \cite{Saida:1999ec,Kraus:2005vz}), thus confirming the validity of the boundary stress tensor.

It is amusing to note that the counterterm in (\ref{regten}) is proportional to the central charge. Thus, at the special point where the central charge vanishes (the so-called chiral point), the divergence of the stress tensor disappears without counterterm, leading directly to a finite result. This is in contrast to the divergences encountered in topologically massive gravity, where a counterterm is required even at the chiral point 
\cite{Kraus:2005zm,Grumiller:2008qz}.

\subsection{BTZ black holes}
The BTZ black hole is a solution of pure AdS gravity that is locally equivalent to AdS$_3$
\cite{Banados:1992wn}. Despite being locally `trivial', it has an event horizon and, in general, a non-vanishing entropy and mass, the latter of which will be computed in the following. The metric is given by
 \bea
  ds^2 \ = \ N^2 dr^2-N^{-2}dt^2+r^2\left(d\phi+N_{\phi}dt\right)^2\;,
 \eea
where
 \bea
  N(r) \ = \ \left(-8GM+\frac{r^2}{\ell^2}+\frac{16 G^2J^2}{r^2}\right)^{-\frac{1}{2}}\;,
  \qquad
  N_{\phi} \ = \ -\frac{4GJ}{r^2}\;,
 \eea
with $|J|\leq M\ell$.
In general there are two event horizons located at
 \bea
  r_{\pm}^2 \ = \ 4GM\ell^2 \left(1\pm\left[1
  -\left(\tfrac{J}{M\ell}\right)^2\right]^{\frac{1}{2}}\right)\;,
 \eea
and we observe that
absence of naked singularities requires $M> 0$.

To compute mass and angular momentum we insert this metric into the regularized stress tensor (\ref{regten}). We obtain from (\ref{massmomentum})
 \bea\label{BTZmass}
  M_{\rm BTZ} \ = \ M\left(\sigma+\frac{1}{2m^2\ell^2}\right)\;, \qquad
  J_{\rm BTZ} \ = \ J\left(\sigma+\frac{1}{2m^2\ell^2}\right)\;,
 \eea
where we used for the spatial boundary coordinate
$\int_{\Sigma}dx\rightarrow \ell\int_{0}^{2\pi}d\phi$. This result agrees with \cite{Clement:2009gq}. We observe that the values are proportional to the central charges (\ref{central}). In particular we conclude that for physical solutions with $M> 0$ the mass of the BTZ black hole is positive whenever the central charge is positive. One of the main results of \cite{Bergshoeff:2009aq} was that the bulk gravitons propagate unitarily and non-tachyonic whenever the central charge is \textit{negative}. So we encounter here the same conflict as  in topologically massive gravity with negative cosmological constant \cite{Li:2008dq}: either one chooses the parameters such that the graviton modes have positive energy (i.e.~are not ghosts), then the BTZ black holes have negative mass, or one chooses the parameters such that the BTZ black holes have positive mass, but then the gravitons are ghosts. Both choices indicate an instability of the AdS vacua. In the case of NMG, however, there is one exception, namely the chiral point corresponding to
$c_{L}=c_{R}=0$. At this point the bulk modes are massive spin-1 excitations that propagate unitarily and non-tachyonic for the `wrong-sign' Einstein-Hilbert term, and the mass of the BTZ black holes is zero for both sign choices, as follows from (\ref{BTZmass}). At the same time, there are new black holes solutions, which are not locally equivalent to AdS$_3$, and we turn now to the important question whether their mass is positive for the same choice of parameters.

\subsection{New black holes at the chiral point}
New black hole solutions that are not locally AdS$_3$ have been given by Clement in \cite{Clement:2009ka}.
The general metric considered in \cite{Clement:2009ka} is given by
 \bea\label{clement}
  ds^2 \ = \ -\left(\frac{2\rho}{\ell^2}-F(\rho)\right)dt^2+\frac{\ell^2}{4\rho^2}d\rho^2
  +\left(2\rho +\ell^2 F(\rho)\right)d\varphi^2-2\ell F(\rho)dtd\varphi\;.
 \eea
This reduces for $F(\rho)=0$ to AdS$_3$, which is related to the metric in the form (\ref{ads3}) via the coordinate transformation $r^2=2\rho$.
Specifically, at the chiral point a new asymptotically AdS solution (in the sense of log gravity) has been found which is defined by
 \bea
  F(\rho) \ = \ d\ln \biggl|\frac{\rho}{\rho_0}\biggr|\;.
 \eea
Inserting this into the stress tensor and then into (\ref{massmomentum}) we obtain for mass and angular momentum
 \bea\label{CBH}
  M \ = \ \frac{2d\sigma}{G}\;, \qquad J \ = \ \frac{2d\ell\sigma}{G}\;,
 \eea
which coincides with the results found in \cite{Clement:2009ka} by other methods.
In \cite{Clement:2009ka} it has been shown that this metric is devoid of naked singularities if $d<0$. From (\ref{CBH}) we infer that the physical mass for this choice is positive only provided $\sigma=-1$. Thus we conclude that at the chiral point all requirements on both  physically acceptable black hole solutions and unitarily propagating bulk modes consistently lead to the `wrong-sign' Einstein-Hilbert term.

\section{Lifshitz backgrounds and black holes}
In this section we are going to discuss to what extent a finite stress tensor can be defined for Lifshitz backgrounds. We first review the appearance of Lifshitz solutions in NMG, slightly extending the discussion of \cite{AyonBeato:2009nh}. We then turn to the black hole solution that is asymptotically Lifshitz given in \cite{AyonBeato:2009nh} and discuss the computation of its mass.

\subsection{Lifshitz vacua and their stress tensor}
We start with a discussion of the appearance of Lifshitz solutions in NMG. The latter are characterized by a different scaling of space and time and have attracted much interest recently due to their possible application as gravity duals to non-relativistic systems as, e.g., in condensed matter physics. Their metric is given by \cite{Kachru:2008yh}
 \bea\label{Lifshitz}
  ds^2 \ = \ \frac{\ell^2}{r^2}dr^2
  -\frac{r^{2z}}{\ell^{2z}}dt^2+\frac{r^2}{\ell^2}dx^2\;,
 \eea
where $z$ is the so-called dynamical exponent. Here, $x$ generically denotes the spatial directions of the `boundary part', which will be one-dimensional in the case we are interested in. The boundary part is invariant under the anisotropic scaling
 \bea
  t \ \rightarrow \ \alpha^z t\;, \qquad x \ \rightarrow \ \alpha x\;,
 \eea
provided $r\rightarrow \alpha^{-1}r$.
This reduces to the standard scale invariance of relativistic CFTs for $z=1$, for which the Lifshitz metric (\ref{Lifshitz}) coincides with the AdS metric (\ref{ads3}).

Inserting the Lifshitz ansatz into the NMG field equations shows that they are solved provided the cosmological parameter in the action is given by
 \bea\label{lambdaz}
  \lambda \ = \ -\frac{z^2+z+1}{z^2-3z+1}\;.
 \eea
Moreover, the Lifshitz length scale $\ell$ is determined by the field equations according to
 \bea\label{ellLif}
  \ell^2 \ = \ \frac{1}{2m^2\sigma}\left(z^2-3z+1\right)\;.
 \eea

Let us now inspect the space of Lifshitz solutions in more detail. The relation
(\ref{lambdaz}) provides a quadratic equation for $z$ and, accordingly, for a given model with fixed $\lambda$ there are generically two Lifshitz vacua. Their dynamical exponents are
 \bea\label{zsol}
  z_{\pm} \ = \ -\frac{1-3\lambda\pm\sqrt{5\lambda^2-14\lambda-3}}{2(1+\lambda)} \;.
 \eea
For maximally symmetric vacua the points $\lambda=-1$ and $\lambda=3$ are special. The former is special in that the two (A)dS vacua coincide, the latter because it corresponds to the `chiral point' at which the central charges of the dual CFT vanish \cite{Bergshoeff:2009aq}. These points are also special in the case of Lifshitz solutions. First, for $\lambda=-1$ there is a unique solution with $z=0$. More precisely, the $z_{-}$ solution is regular at $\lambda=-1$, despite appearance, where it is a zero, while $z_{+}$ has a singularity. Second, we infer from (\ref{zsol}) that the solutions are real except in the interval $\lambda\in(-\tfrac{1}{5},3)$. Thus, there is a `gap' in the space of Lifshitz solutions after $\lambda=-\tfrac{1}{5}$ (corresponding to another special point $P_1$), which restart again at $\lambda=3$, i.e., at the second special point $P_2$. Precisely at
$\lambda=3$ we have $z=1$, i.e., the Lifshitz metric reduces to the AdS metric. See the figure.

\begin{figure}[t]
\begin{center}
\includegraphics[width=13.4cm]{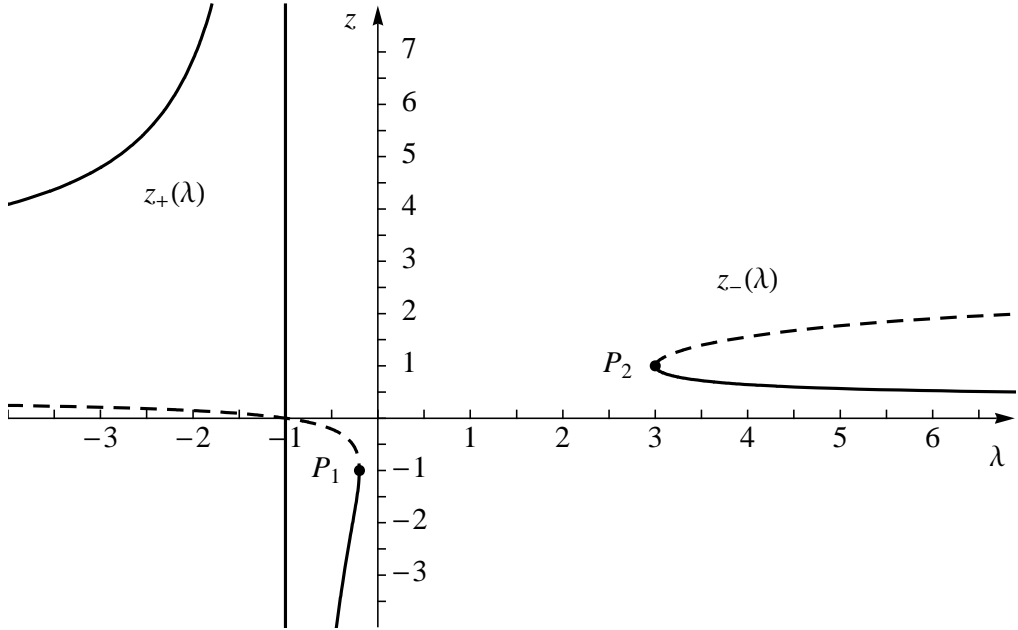}
\end{center}
\caption{Lifshitz vacua corresponding to $z_{+}(\lambda)$, indicated by a solid line, and 
$z_{-}(\lambda)$, indicated by a dashed line, see (\ref{zsol}). $P_1$ and $P_2$ are the special points mentioned in the text. 
\label{plot lifshitz}}
\end{figure}

We turn now to the question whether a renormalized stress tensor can be defined. Evaluating the Brown-York tensor for the Lifshitz solution (\ref{Lifshitz}), we find for the time component
 \bea
  8\pi G\, T_{00} \ = \ -\frac{r^{2z}}{\ell^{2z+1}}\left(\sigma-\frac{z^2-3z+1}{2m^2\ell^2}
  \right)\;.
 \eea
As before, here we have inserted the on-shell values for the auxiliary fields in the Lifshitz background, which implies in particular
 \bea\label{auxlif}
  \hat{s} \ = \ -\frac{z^2-z+1}{m^2\ell^2}\;, \qquad
  \hat{f} \ = \ -\frac{2z}{m^2\ell^2}\;.
 \eea
Remarkably, using the relation (\ref{ellLif}) that is satisfied for the Lifshitz solutions in NMG, we find $T_{00}= 0$.
Thus, while generally the mass computed from (\ref{massform}),
 \bea\label{MassLif}
  M \ = \ \int_{\Sigma}dx\,\frac{\ell^{z-1}}{r^{z-1}}\,T_{00}
  \ = \ -\frac{1}{8\pi G\,\ell^{z+2}}\left(\sigma-\frac{z^2-3z+1}{2m^2\ell^2}\right)\int
  dx\;r^{z+1} \longrightarrow \infty\;,
 \eea
diverges once we approach the boundary at $r\rightarrow\infty$, which could again be renormalized by adding a counterterm $\int d^2x\sqrt{-\gamma}$, for the particular class of Lifshitz solutions they are already finite. One can check that this holds also for
$T_{01}=T_{10}$ and $T_{11}$. In this sense they resemble the AdS case at the chiral point.
However, this does not exclude the possibility that a counterterm is required in order to obtain finite results for other solutions. Evidently, such a counterterm would have to vanish when specialized to the Lifshitz vacua. We will see next that precisely this happens for the Lifshitz black hole.

\subsection{Asymptotically Lifshitz black holes}
We turn now to the particular point where a black hole has been found that is asymptotically Lifshitz \cite{AyonBeato:2009nh}. This solution might prove to be important for finite temperature effects in potential applications as the gravity dual of condensed matter systems. The metric ansatz reads
 \bea\label{LifBH}
  ds^2 \ = \ \frac{\ell^2}{r^2}H(r)dr^2-\frac{r^{2z}}{\ell^{2z}}F(r)dt^2
  +\frac{r^2}{\ell^2}dx^2\;.
 \eea
The functions are given by
 \bea
  F(r) \ = \ H^{-1}(r) \ = \ 1-\frac{r_{H}^2}{r^2} \;,
 \eea
where $r_{H}$ denotes a free integration constant. Analogous to the Schwarzschild solution
in $D=4$, this metric has a curvature singularity at $r=0$ and a single event horizon
at $r=r_{H}$. Inserting (\ref{LifBH}) into the NMG equations and using (\ref{lambdaz})
and (\ref{ellLif}) shows that the only Lifshitz solutions have $z=3$. We then infer that
 \bea\label{BHloc}
  \lambda \ = \ -13\;, \qquad 2m^2\ell ^2 \ = \ \sigma\;.
 \eea
From the second equation of (\ref{BHloc}) we conclude that $m^2$ has to be chosen to be negative for $\sigma=-1$, which case is contained in our general analysis so far.

Next, we discuss the computation of the mass. Inserting the black hole solution (\ref{LifBH}) into the stress tensor we obtain
 \bea\label{Bhdiv}
   T_{00} \ = \ \frac{
   r_H^2}{2\pi G \ell^7\sigma}r^4+{\cal O}(r^2)\;,
 \eea
which gives rise to a divergence as $r \rightarrow \infty$ upon insertion into the mass formula given in (\ref{MassLif}). This cannot be renormalized by adding only a cosmological constant term, because this would upset the finiteness for pure Lifshitz vacua found above. However, in the Lifshitz case we have more possibilities for adding a local counterterm. For instance, we can add terms involving any powers of $\hat{f}$
and $\hat{s}$ because these are scalars under boundary-preserving diffeomorphisms. However, given the fact that on-shell these are related to the (bulk) curvature tensor and that we are dealing with theories quadratic in the curvature, let us illustrate the procedure with  counterterms quadratic in $\hat{f}$ or $\hat{s}$. Their on-shell values for this black hole solution are given by 
 \bea
  \hat{s} \ = \ -\frac{7}{m^2\ell^2}\;, \qquad 
  \hat{f} \ = \ -\frac{2}{m^2\ell^2}\left(3-\frac{2r_{H}^2}{r^2}\right)\;,
 \eea
and therefore only $\hat{f}$ is sufficient in order to cancel the divergence.
Thus, we make the following ansatz for the action including boundary counterterms
 \bea
  S_{\rm ren} \ = \ S_{\rm NMG}+S_{\rm GGH}+\frac{1}{16\pi G}\int d^2x\,
  \sqrt{-\gamma}\,(\alpha_1+\alpha_2\hat{f}+\alpha_3\hat{f}^2)\;,
 \eea
which gives rise to the renormalized stress tensor
 \bea\label{lifstress}
  T_{ij}^{\rm ren} \ = \ T_{ij}+\frac{1}{16\pi G}
  (\alpha_1+\alpha_2\hat{f}+\alpha_3\hat{f}^2)\gamma_{ij}\;.
 \eea
Here we have three free coefficients, but only two conditions to fix them. First, the counterterm has to vanish when evaluated for pure Lifshitz solutions in order to be consistent with their finiteness without counterterm. Second, the counterterm has to cancel the divergence in (\ref{Bhdiv}). Thus, in presence of only a single Lifshitz black hole solution, the counterterm is not uniquely determined. 
We conclude that at present our understanding of the Lifshitz case is insufficient in order 
to determine the correct boundary stress tensor unambiguously. One might, however, speculate 
that a third condition can be obtained by assuming the standard thermodynamic relation 
$dM=TdS$. If one assumes that the entropy can be computed via 
Wald's formula, whose validity in the Lifshitz case is in fact is not clear to us, one can apply the result of \cite{Cai:2009ac}.  (For another approach towards the thermodynamics of Lifshitz black holes we refer to \cite{Myung:2009up}.) Given the (positive)  
Hawking temperature as determined in \cite{AyonBeato:2009nh},
this implies for the mass 
 \bea\label{lifshitzmass}
  M \ = \ -\frac{\sigma}{4G}\left(\frac{r_{H}}{\ell}\right)^4\;.
 \eea
The requirement that this value follows from the stress tensor (\ref{lifstress}) finally fixes all coefficients,
 \bea
  \alpha_1 \ = \ 
  \frac{15\sigma}{2\ell}\;, \qquad
  \alpha_2 \ = \ \frac{1}{4\ell}\;, \qquad
  \alpha_3 \ = \ -\frac{\sigma}{32\ell}\;.
 \eea
Unfortunately, given the ambiguity of the counterterm, this computation does not provide an independent check for the mass. Rather this has to await the knowledge of more Lifshitz black hole solutions or a convincing physical principle that might fix the form of the counterterm.

\section{Conclusions and discussion}
In this paper we investigated the definition of the Brown-York stress tensor for curvature-squared theories in generic dimensions, with a special emphasis on new massive gravity. This requires a formulation of the gravitational theory that admits a well-posed variational principle. While for generic curvature-squared theories in the usual formulation there is no Gibbons-Hawking like term that is sufficient, we provided here an on-shell equivalent reformulation involving an auxiliary tensor field carrying the symmetries of the Riemann tensor that does allow for a Gibbons-Hawking term. It is remarkable that the latter takes the simple form (\ref{gengenGH}), which is completely universal in that it does not depend on the details of the theory. This result might be useful for other applications as well. 

Given this reformulation, the boundary stress tensor is well-defined according to 
(\ref{stressten0}). We provided evidence that this prescription involving the auxiliary field is reasonable by computing physical quantities for new massive gravity. In particular, we confirmed the values of the central charges and the mass of the BTZ black holes. In addition, we computed the mass of black holes that are asymptotically AdS (but, in contrast to BTZ black holes, not isometric to AdS$_3$), and we discussed the asymptotically Lifshitz case. For the latter we proposed a novel covariant counterterm, which leads to a finite result for the black hole solution. It remains, however, to be seen whether this is the right prescription and whether the analogous procedure works also in higher dimensions. In particular, it would be 
desirable to have a reasoning from first principles that determines the counterterm uniquely.  
In this context we mention that several new Lifshitz black hole solutions of higher-derivative gravity have been found in $D\geq 4$ while this paper was in the stages of completion \cite{AyonBeato:2010tm}, and it would be interesting to attempt the computation of mass and other physical quantities for these solutions by the method outlined here.

For the asymptotically AdS black hole we found that its mass is physical if and only if one chooses the `wrong-sign' Einstein-Hilbert term. For the special case of `chiral new massive gravity' this implies that all physical requirements, namely that both the gravitons and black holes have positive mass/energy, lead to a consistent choice for the sign of the Einstein-Hilbert term. (This is in contrast to topologically massive gravity at the chiral point which still exhibits a conflict between gravitons and BTZ black holes.) Thus we provided further evidence that chiral new massive gravity might be `well-behaved' precisely for the `wrong-sign' Einstein-Hilbert term, see also \cite{Grumiller:2009sn}. On the other hand, for the Lifshitz case there is so far no independent check for the sign of the Einstein-Hilbert term, for the behavior of graviton modes about the Lifshitz vacua has not been investigated. We leave this for future research.

The computation of the stress tensor provides just the first step of the AdS/CFT program. Further results for chiral new massive gravity have already appeared in \cite{Grumiller:2009sn}, where it has been shown at the level of 2-point functions that the dual CFT is a logarithmic CFT. This research can be extended into various directions. For instance, it would be interesting to examine curvature-squared theories, in three and higher dimensions, in regard to the question whether the field theories dual to any of these models is of relevance, say, in condensed matter physics.

\subsection*{Acknowledgments}

We acknowledge helpful discussions with Allan Adams, Eric Bergshoeff, Gaston Giribet, Daniel Grumiller, Roman Jackiw, Hong Liu, John McGreevy, Pietro Menotti, Paul Townsend and 
Mattias Wohlfarth.  

This work is in part supported by funds
provided by the U.S. Department of Energy (DoE) under the cooperative
research agreement DE-FG02-05ER41360.
The work of OH is supported by the DFG -- The German Science Foundation. 
The work of ET is supported by INFN through a Bruno Rossi fellowship.

\begin{appendix}
\renewcommand{\theequation}{\Alph{section}.\arabic{equation}}

\section{ADM variables and boundary diffeomorphisms} \setcounter{equation}{0}
In this appendix we present further details on the ADM decomposition used for the computation of the boundary variation and the corresponding counterterms. In terms of ADM variables involving lapse and shift functions $N$ and $N_i$, the bulk metric is parameterized as
 \bea\label{ADMmetric}
  g_{\mu\nu} \ = \
  \left(\begin{array}{cc} \gamma_{ij} &
  N_i \\ N_j & N^2+N^{i}N_{i} \end{array}\right)\;,
 \eea
where $\gamma_{ij}$ denotes the boundary metric. A similar decomposition is used for the auxiliary field $f^{\mu\nu}$, see eq.~(\ref{fdeco}).

We now inspect the presence of gauge symmetries in this decomposition. Any generally covariant theory is invariant under diffeomorphisms whose infinitesimal action on the fields is given by the Lie derivative with respect to a vectorial parameter $X^{\mu}$, i.e.,
 \bea\label{diff}
  \delta_{X}g_{\mu\nu} &=& {\cal L}_{X}g_{\mu\nu} \ \equiv \ X^{\rho}\partial_{\rho}g_{\mu\nu}
  +\partial_{\mu}X^{\rho}g_{\rho\nu}+\partial_{\nu}X^{\rho}g_{\mu\rho}\;, \\ \nonumber
  \delta_{X}f^{\mu\nu} &=& {\cal L}_{X}f^{\mu\nu} \ \equiv \ X^{\rho}\partial_{\rho}f^{\mu\nu}
  -f^{\mu\rho}\partial_{\rho}X^{\nu}-f^{\rho\nu}\partial_{\rho}X^{\mu}\;.
 \eea
In the ADM decomposition, the vectorial diffeomorphism parameter can be decomposed accordingly,
 \bea
  X^{\mu} \ = \ \left( \xi^{i},\lambda \right)\;,
 \eea
where a priori $\xi^i$ and $\lambda$ are arbitrary functions of $x^{\mu}=(x^i,r)$.  However, once the coordinates are adapted to the boundary in the sense that the normal vector
on the boundary in this coordinates is given by $n_{\mu}=(0,0,N)$ (compare eq.~(\ref{covvar})), the diffeomorphism symmetry is reduced. Acting with a general diffeomorphism on $n_{\mu}$ implies
 \bea\label{deltan}
  \delta n_{\mu} \ = \ X^{\rho}\partial_{\rho}n_{\mu}+\partial_{\mu}X^{\rho}n_{\rho}
  \qquad \Rightarrow\qquad \delta n_i \ = \ \partial_{i}\lambda\,N\;.
 \eea
Thus, in order to maintain $n_i=0$, the diffeomorphism parameter along the radial direction is restricted to be independent of the $x^i$, i.e., $\lambda=\lambda(r)$. We refer to these as `boundary-preserving diffeomorphisms'.

Next, let us evaluate the  boundary-preserving diffeomorphism symmetry for the ADM variables. Applying (\ref{diff}) to (\ref{ADMmetric}) and setting $\partial_i\lambda=0$, we obtain
 \bea\nonumber
  \delta \gamma_{ij} &=& {\cal L}_{\xi}\gamma_{ij}+\lambda\partial_{r}\gamma_{ij}\;,
  \\ 
  \delta N_i &=& {\cal L}_{\xi}N_i+\lambda \partial_rN_i+\partial_r\lambda\, N_i
  +\partial_r\xi^{j}\,\gamma_{ij}\;, \\ \nonumber
  \delta N &=& {\cal L}_{\xi}N+\lambda\partial_r N+\partial_r\lambda\, N\;.
 \eea
In particular, one may verify that the transformation rule for $N$ under these diffeomorphisms is consistent with its transformation as the $r$-component of $n_{\mu}$ according to (\ref{deltan}). With these transformation rules at hand we are now able to derive the rules of gauge-fixing and undoing the gauge-fixing as used in the main text. After choosing Gaussian normal coordinates with $N=1$ and $N_i=0$, the residual diffeomorphisms are parametrized by $X^{\mu}=(\xi^i (x^i),\lambda)$, where $\lambda$ is now a constant. Under this residual symmetry $\partial_r \gamma_{ij}$ transforms covariantly in the sense that it transforms in the same way as $\gamma_{ij}$. This is not the case in the most general formulation without gauge-fixing, and so in this case $\partial_r \gamma_{ij}$ has to be suitably `covariantized'. As to be expected, the `covariant $r$-derivative' is given by the extrinsic curvature (\ref{extrinsic}) in that the latter transforms as
 \bea
  \delta K_{ij} \ = \ {\cal L}_{\xi}K_{ij}+\lambda \partial_r K_{ij}\;.
 \eea
Thus, gauge invariance shows that in order to undo the gauge-fixing we have to replace $-\tfrac{1}{2}\partial_r\gamma_{ij}$ by the full $K_{ij}$.

Let us now repeat this analysis for $f^{\mu\nu}$. Under the boundary-preserving diffeomorphisms its components transform according to (\ref{diff}) as
 \bea\nonumber
  \delta f^{ij} &=& {\cal L}_{\xi}f^{ij}+\lambda\partial_r f^{ij}-2h^{(i}\partial_r\xi^{j)}
  \;, \\
  \delta h^i &=& {\cal L}_{\xi}h^i +\lambda\partial_r h^i-h^i\partial_r\lambda-s\partial_r
  \xi^i\;, \\ \nonumber
  \delta s &=& {\cal L}_{\xi}s+\lambda\partial_r s-2s\partial_r\lambda\;.
 \eea
We observe that the transformation rules receive non-covariant terms which vanish only for the residual diffeomorphisms after gauge-fixing to normal coordinates. In order to remedy this, we introduce combinations involving lapse and shift functions in such a way that they transform covariantly in general. Specifically, the hatted quantities (\ref{hatted}) transform as
 \bea
  \delta_X \hat{f}^{ij} &=& {\cal L}_{\xi}\hat{f}^{ij}+\lambda\partial_r \hat{f}^{ij}\;,
  \qquad \text{etc.}\;,
 \eea
as required. Also, we have to introduce `covariant $r$-derivatives',
 \bea
  {\cal D}_r \hat{f}^{ij} &=& \frac{1}{N}\left(\partial_r\hat{f}^{ij}-
  N^k\partial_k\hat{f}^{ij}+\hat{f}^{kj}\partial_k N^i+\hat{f}^{ik}\partial_k N^j\right)\;, \\
  {\cal D}_r\hat{f} &=& \frac{1}{N}\left(\partial_r\hat{f}-N^j\partial_j \hat{f}\right)\;,
 \eea
which transform in the same way as $\hat{f}^{ij}$ and $\hat{f}$.

The previous discussion can actually be casted into a more covariant form, which we briefly review in the following. Instead of splitting the
indices into boundary and normal directions, one can formally maintain the full covariance by introducing projection operators. For this we
use the normal vector $n_{\mu}$, whose contravariant components read
 \bea
  n^{\mu} \ = \ \left(-N^{-1}N^{i},N^{-1}\right)\;.
 \eea
The boundary metric can then be defined `covariantly' as
 \bea
  \gamma^{\mu\nu} \ = \ g^{\mu\nu}-n^{\mu}n^{\nu} \ = \ \left(\begin{array}{cc} \gamma^{ij} &
  0 \\ 0 & 0 \end{array}\right)\;,
 \eea
where we stress that the last equality holds without gauge-fixing. Upon raising and lowering indices with $g_{\mu\nu}$ one obtains a projector
from the tensor with mixed indices,
 \bea
  \gamma_{\mu}{}^{\nu} \ = \ \delta_{\mu}{}^{\nu}-n_{\mu}n^{\nu} \ = \ \left(\begin{array}{cc} \delta_{i}{}^{j} &
  0 \\ N^{j} & 0 \end{array}\right)\;.
 \eea
We should point out that this covariant language is somewhat formal in that the vector $n_{\mu}$ reflecting the presence of the
boundary still transforms only under the subgroup of boundary-preserving diffeomorphisms as a vector.
Nevertheless, the covariant combinations (\ref{hatted}) of the auxiliary fields can now simply be written as
 \bea
  \hat{f}^{ij} &=& f^{\mu\nu}\gamma_{\mu}{}^{i}\gamma_{\nu}{}^{j}\;, \\
  \hat{h}^{i} &=& f^{\mu\nu}\gamma_{\mu}{}^{i}n_{\nu}\;, \\
  \hat{s} &=& f^{\mu\nu}n_{\mu}n_{\nu}\;.
 \eea
Analogously, one finds for the auxiliary field corresponding to the full Riemann tensor
 \bea
  \hat{\phi}^{ij} &=&  \phi^{\mu\nu\rho\sigma}\gamma_{\mu}{}^{i}\gamma_{\rho}{}^{j}n_{\nu}
   n_{\sigma} \ = \ N^2\phi^{irjr}\;, \\
  \hat{\phi}^{ijk} &=& \phi^{\mu\nu\rho\sigma}n_{\mu}\gamma_{\nu}{}^{i}\gamma_{\rho}{}^{j}
  \gamma_{\sigma}{}^{k} \ = \ 
  N\left(\phi^{rijk}-2\phi^{rir[j}N^{k]}\right)\;, \\
  \hat{\phi}^{ijkl} &=& \phi^{\mu\nu\rho\sigma}\gamma_{\mu}{}^{i}\gamma_{\nu}{}^{j}
  \gamma_{\rho}{}^{k}\gamma_{\sigma}{}^{l} \\ \nonumber
  &=& \phi^{ijkl}-2N^{[i}\phi^{j]rkl}-2\phi^{ijr[k}N^{l]}
  +4N^{[i}\phi^{j]rr[k}N^{l]}\;, 
 \eea
where we used the Riemann-tensor symmetries of $\phi^{\mu\nu\rho\sigma}$.

\section{Auxiliary fields in arbitrary dimensions} 
Here we present for completeness some relations for the auxiliary-field formulation in arbitrary dimensions. 
The auxiliary field has the same algebraic symmetries as the Riemann tensor, corresponding to the (2,2) Young tableau, 
 \bea\label{22Young}
  \phi_{\mu\nu\rho\sigma} \ = \ -\phi_{\nu\mu\rho\sigma}\;, \qquad
  \phi_{\mu\nu\rho\sigma} \ = \ -\phi_{\mu\nu\sigma\rho}\;, \qquad 
  \phi_{[\mu\nu\rho]\sigma} \ = \ 0\;,
 \eea
which imply also the `exchange property' $\phi_{\mu\nu\rho\sigma}=\phi_{\rho\sigma\mu\nu}$. 
Taking this symmetry into account, the field equations for $\phi$ read
\begin{equation}\label{phieq}
R_{\mu\nu\rho\sigma}+2b_1 \phi_{\mu\nu\rho\sigma}
+2b_2 \phi_{\langle \mu\rho} g_{\nu\sigma\rangle}
+2b_3 \phi\,g_{\langle \mu\rho} g_{\nu\sigma\rangle}
\,=\,0\;,
\end{equation}
where 
$\langle\; \rangle$ denotes the projector which imposes the symmetries (\ref{22Young}).
Explicitly, we have, e.g., 
 \bea
  \phi_{\langle\mu\rho}g_{\nu\sigma\rangle} = 
  \frac{1}{4} \left(\phi_{\mu\rho}g_{\nu\sigma}-\phi_{\nu\rho}g_{\mu\sigma}
  -\phi_{\mu\sigma}g_{\nu\rho}+\phi_{\nu\sigma}g_{\mu\rho}\right)\;.
 \eea
Taking the single and double trace of (\ref{phieq}), one obtains the relations   
\begin{eqnarray}
\phi &=& -\,\frac{R}{2 b_1+d b_2+d(d+1)b_3}\;,
\\
\rule{0pt}{.7cm}
\phi_{\mu\nu} &=&
-\, \frac{2}{4 b_1+(d-1)b_2} \left( R_{\mu\nu}
- \frac{b_2+2d b_3}{2[2 b_1+d b_2+d(d+1)b_3]}\, R \,g_{\mu\nu}\right)\;,
\\
\rule{0pt}{.7cm}
\phi_{\mu\nu\rho\sigma} &=&
-\,\frac{1}{2b_1}\left\{
R_{\mu\nu\rho\sigma}- \frac{4 b_2}{4 b_1+(d-1)b_2} R_{\langle \mu\rho} g_{\nu\sigma\rangle}\right.\\
\rule{0pt}{.7cm}& & \hspace{1.5cm}
\left.+\,\frac{2[b_2^2-4 b_1 b_3+(d+1)b_2 b_3]}{[4 b_1+(d-1)b_2][2 b_1+d b_2+d(d+1)b_3]}\,
R\, g_{\langle \mu\rho} g_{\nu\sigma\rangle}
\right\}\;.
\nonumber
\end{eqnarray}
Finally, reinserting these on-shell values into (\ref{gensquare}), one obtains the following  relation between the coefficients of (\ref{gensquare}) and    
(\ref{lagrangian riemann squared}),
\begin{eqnarray}\label{a1}
a_1 &=& -\frac{1}{4 b_1}\;,
\\
\rule{0pt}{.6cm}
a_2 &=&
\frac{b_2}{\big[ 4 b_1+(d-1)b_2\big] b_1}\;,
\\
\rule{0pt}{.6cm}
a_3 &=&
-\,\frac{b_2^2-4 b_1 b_3 +(d+1)b_2 b_3}{2b_1 \big[4 b_1+(d-1)b_2\big]\big[ 2b_1+d b_2+d(d+1)b_3\big]}\;.
\end{eqnarray}
We note that the equivalence of (\ref{lagrangian riemann squared}) and (\ref{gensquare}) 
does not hold for all choices of coefficients, because 
$a_1=0$ corresponding to the absence of the square of the Riemann tensor is not contained for $D>3$, as one sees from (\ref{a1}). In this case it is sufficient to work with an auxiliary symmetric 2-tensor as in $D=3$.

\end{appendix}

\end{document}